\documentclass{anstrans}
\title{Functional Expansion Tallies of Matrix Operators for Prediction for Integrated Autocorrelation Time\\
       in Batch Monte Carlo: an Analytic 2D Scattering Chain Benchmark}
\author{Jilang Miao$^{*}$, Guillaume L. Giudicelli$^{\dagger}$, William Reed Kendrick$^{\ddagger}$}

\institute{%
$^{*}$Ken and Mary Alice Lindquist Department of Nuclear Engineering,
Pennsylvania State University, University Park, PA 16802 \\
$^{\dagger}$Idaho National Laboratory, Idaho Falls, ID 83415 \\
$^{\ddagger}$Department of Nuclear Science and Engineering,
Massachusetts Institute of Technology, Cambridge, MA 02139
}

\usepackage{graphicx}
\usepackage{booktabs}
\usepackage{multirow}
\usepackage{microtype}
\usepackage{amsmath}
\usepackage{url}

\newcommand{\tauint}{\tau_{\mathrm{int}}}
\newcommand{\lam}{\lambda}
\newcommand{\rb}{r_b}

\begin{document}
\maketitle
{\let\thefootnote\relax\footnotetext{$^\S$Corresponding author: Jilang Miao, \texttt{jlmiao@psu.edu}}}

%%%%%%%%%%%%%%%%%%%%%%%%%%%%%%%%%%%%%%%%%%%%%%%%%%%%%%%%%%%%%%%%%%%%%
\section{Introduction}

Functional expansion tallies (FETs) are a successful example of using reduced
bases as Monte Carlo (MC) tally estimators rather than only as post-processing % G edit
representations.  Instead of estimating many mesh-bin values independently, an
FET estimates the coefficients of a flux, reaction-rate, or source distribution,
usually in an orthogonal basis, and then reconstructs the continuous distribution from % G edit: we are doing NOFETs at INL
those coefficients~\cite{Griesheimer2006,Herman2014}.  For sufficiently smooth
distributions, this can converge faster and with less residual error than a
histogram tally.  In representative RMC studies, FET reconstructions were
consistent with mesh tallies while reducing memory and runtime; order-selection
strategies based on coefficient uncertainty were also proposed~\cite{Wang2021}.
In multiphysics applications, these coefficients also provide a compact
data-transfer representation between MC and finite-element solvers, reducing % cite Matt thesis?
dependence on a common fine mesh~\cite{Ellis2017}.  This coefficient-space view is the starting
point of the present work.  If a Monte Carlo simulation is already tallying
expansion coefficients for FET, then it is natural to ask whether the
generation-to-generation correlation operator can also be tallied in the same
reduced basis.

That question matters because neutron generations in MC criticality calculations are not independent. % G edit
The fission source evolves as a Markov chain, and
inter-cycle correlations inflate tally variances~\cite{Gelbard1990}.
Let $F = \int_I f(\mathbf{r})\,\pi(d\mathbf{r})$ be a tally  of event f over a 
filter region $I$ (in this work, a spatial subdomain $I\subset[-L,L]^2$),
and let $\rho_I(d)$ denote its inter-cycle autocorrelation at generation lag $d$.
If $\hat{F}$ is the sample mean of the tally over $N$ batches, its true variance is
\begin{equation}\label{eq:var}
  \mathrm{Var}(\hat{F}) = \frac{\mathrm{Var}(F)}{N}
    \left[1 + 2\sum_{d=1}^{N-1}\!\Bigl(1-\tfrac{d}{N}\Bigr)\rho_I(d)\right],
\end{equation}
For $N$ well above the correlation length the $d/N$ term is negligible,
and the bracket converges to the
\emph{integrated autocorrelation time} (IACT)
$\tauint \equiv 1 + 2\sum_{d=1}^{\infty}\rho_I(d) \ge 1$,
so that $\mathrm{Var}(\hat{F})\approx\tauint\cdot\mathrm{Var}(F)/N$.
This paper works in the large-$N$ regime where $\tauint$ is sufficient.
Neglecting correlation therefore underestimates the standard error by
$\sqrt{\tauint}$, which can exceed 3--10 for large, slowly mixing
systems~\cite{Ueki2002,Herman2014,Miao2016}.  Correlation prediction is not only a post-processing
uncertainty correction.  The same information identifies slow source modes,
problematic tally regions, and source-update or variance-reduction strategies
whose benefit depends on the correlation structure~\cite{Miao2020}.

Several reduced or surrogate models have been proposed for this purpose.
Demaret et al.\ fitted AR/MA time-series models for eigenvalue calculations
~\cite{Demaret1999}.  Yamamoto, Sakata, and Endo later expanded the fission
source in diffusion-equation modes and modeled the expansion coefficients with
an autoregressive process to predict variance underestimation in fission-rate
tallies~\cite{Yamamoto2013,Yamamoto2014}.  This is an important predecessor to
the present reduced-basis viewpoint, but it relies on a diffusion surrogate for
the transport problem.  Discrete-state and Multitype Branching Process (MBP)
models improve the physical description and can predict correlation coefficients
and variance-underestimation factors accurately~\cite{Sutton2017,Miao2018}.
Their limitation is dimensional: as the phase-space mesh is refined, the moment
and transition matrices become too large for routine use.

This work tests whether a functional-expansion tally / model-order-reduction
(FET/MOR) representation can remove that discrete-state
bottleneck while retaining the correlation-prediction capability.  The expensive
part of the calculation is still the MC random walk; the reduced operator is an
additional FET-like tally accumulated during that walk.  Thus, in a production
setting where FET coefficients are already assembled for spatial tallies or
source diagnostics, the shared transport cost should not be charged uniquely to
the correlation model.  The present paper isolates the operator-approximation
question in a benchmark with an exact answer: a 2D isotropic scattering chain
with reflective boundaries.  We compare:

\textbf{1. Discrete $M$-cell chain}: tally a dense $M{\times}M$
        transition count matrix from the MC walk; estimate $\tauint$ via
        a truncated Neumann series.

\textbf{2. Galerkin MOR}: project the transition kernel onto an
        $\rb$-dimensional 2D basis (cosine, Legendre, or Chebyshev
        products); accumulate small Galerkin matrices via graphics processing unit (GPU) outer
        products; solve for $\tauint$ via a Karush–Kuhn–Tucker (KKT) constrained resolvent.

Here $M$ is the total cell count ($\sqrt{M}{\times}\sqrt{M}$ grid) and
$\rb$ is the total number of basis pairs ($\sqrt{\rb}{\times}\sqrt{\rb}$
tensor product).

%%%%%%%%%%%%%%%%%%%%%%%%%%%%%%%%%%%%%%%%%%%%%%%%%%%%%%%%%%%%%%%%%%%%%
\section{Theory}

\subsection{2D Scattering Chain}

Consider an isotropically scattering medium on $[-L,L]^2$ with reflective boundaries. %edit G
One generation consists of a single scattering event: starting at $\mathbf{r}$, a neutron travels
distance $s\sim\mathrm{Exp}(\Sigma_t)$ in direction $\theta\sim\mathrm{Uniform}[0,2\pi)$
with reflections at all walls.
The one-generation transition kernel $M^{(1)}(\mathbf{r}{\to}\mathbf{r}')$
has been derived analytically in 1D~\cite{PhdThesisJMiao,Miao2019};
the present work extends those results to 2D (full derivation skipped in this summary).
The full eigenbasis consists of four families (cosine$\times$cosine, cosine$\times$sine,
sine$\times$cosine, and sine$\times$sine); for the symmetric tally $I=[-a,a]^2$ the
sine-containing families integrate to zero and do not contribute to $\tauint$.
The cosine-product eigenfunctions
$\tilde\phi_{mn}(\mathbf{r})=\tilde\phi_m(x)\tilde\phi_n(y)$, where
$\tilde\phi_0{=}1$ and $\tilde\phi_m(x)=\sqrt{2}\cos(m\pi x/L)$ ($m\ge1$),
carry all the weight, with exact eigenvalues
\begin{equation}\label{eq:lam}
  \lam_{mn} = \frac{1}{\displaystyle\sqrt{1 + (m^2+n^2)\Bigl(\frac{\pi}{\Sigma_t L}\Bigr)^2}},
  \quad m,n=0,1,2,\ldots
\end{equation}

\subsection{Exact Integrated Autocorrelation Time} % edit G, best to avoid acronyms in titles

For the square symmetric tally $I=[-a,a]^2$ with $a=\alpha L$, the
$L^2(\pi)$-projection of the centered tally indicator onto eigenmode $(m,n)$
decomposes as $T_{mn}=T_m T_n$, where
\begin{equation}\label{eq:W}
  T_0 = \alpha,\qquad
  T_m = \frac{\sqrt{2}\,\sin(m\pi\alpha)}{m\pi}\;\;(m\ge1).
\end{equation}
From the eigendecomposition of the transition kernel, the generation-lag autocorrelation coefficient at lag $d$ is shown to be  % is this by defintion? if so mention it
\begin{equation}\label{eq:rho}
  \rho_I(d) = \frac{\displaystyle\sum_{(m,n)\neq(0,0)} T_{mn}^2\,\lam_{mn}^d}
                   {\displaystyle\sum_{(m,n)\neq(0,0)} T_{mn}^2},
\end{equation}
and evaluating the $N{\to}\infty$ limit of the bracket in Eq.~\eqref{eq:var}
gives the closed form,
\begin{equation}\label{eq:tau}
  \tauint = 1 + 2\sum_{d=1}^\infty \rho_I(d)
          = 1 + 2\,\frac{\displaystyle\sum_{(m,n)\neq(0,0)} T_{mn}^2\,
            \dfrac{\lam_{mn}}{1-\lam_{mn}}}
                        {\displaystyle\sum_{(m,n)\neq(0,0)} T_{mn}^2}.
\end{equation}

\begin{figure}[ht]
  \centering
  \includegraphics[width=\columnwidth] 
  %0.80\textwidth
  {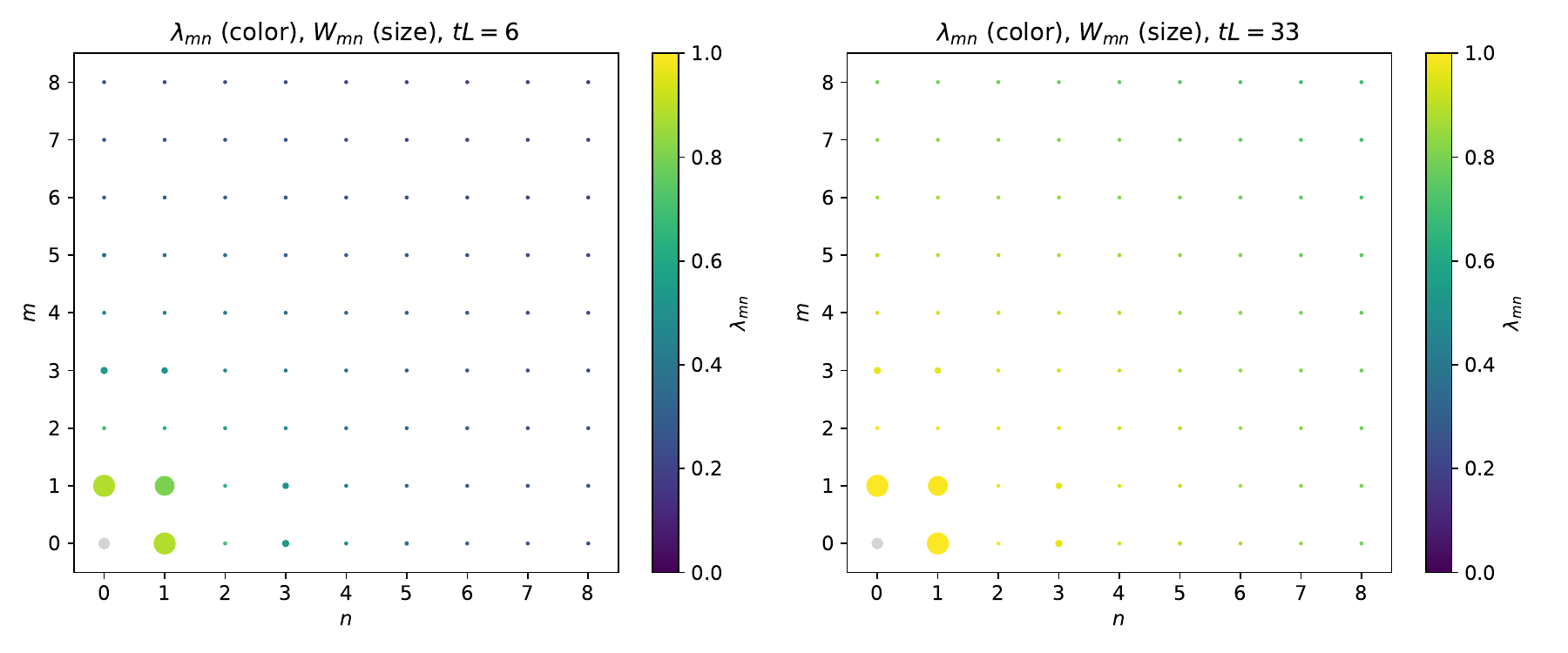}
  \caption{2D mode structure at $\Sigma_t L{=}6$ (left) and $\Sigma_t L{=}33$ (right, used here).
  Each dot at $(m,n)$ encodes eigenvalue $\lam_{mn}$ by color (viridis: dark=large $\lam$,
  bright=small $\lam$) and tally weight $T_{mn}^2$ by marker area.
  At $\Sigma_t L{=}33$ the dominant slow modes cluster near $(0,1)$, $(1,0)$, $(1,1)$.
  Modes with equal $m^2{+}n^2$ share $\lam_{mn}$ (degeneracy); including them step by step
  produces the staircase in the cosine MOR convergence (Fig.~\ref{fig:mor}).}
  \label{fig:modes}
\end{figure}

Fig.~\ref{fig:modes} shows the eigenvalue $\lam_{mn}$ and tally weight $T_{mn}^2$ for each mode at $\Sigma_t L{=}6$ and $33$.

\subsection{Discrete $M$-Cell Approximation}

The domain is partitioned into $M$ equal cells ($\sqrt{M}{\times}\sqrt{M}$
grid).
From the MC walk, transition counts are tallied cell-to-cell,
normalized to a row-stochastic dense
$M{\times}M$ matrix $P$, and $\tauint$ is obtained from the
Neumann series
\begin{equation}\label{eq:neumann}
  \tauint \approx 1 + 2\,\mathbf{h}^\top\sum_{d=1}^{K} P^d\,\mathbf{h}
                  \Big/\mathbf{h}^\top\mathbf{h},
\end{equation}
where $\mathbf{h}$ is the centered tally vector (orthogonal to $\boldsymbol\pi$,
eliminating the $\lam{=}1$ mode).
The dense storage and $O(M^2)$ Neumann cost reflect the full fission-matrix
representation required for this method.
At finite $M$ the chain gives noise-free reference values
$\rho_I(d)=(P^d_{I,I}-\pi_I)/(1-\pi_I)$~\cite{PhdThesisJMiao}, and the exact
eigendecomposition (Eqs.~\eqref{eq:W}--\eqref{eq:tau}) is used for validation.

\subsection{Galerkin Model Order Reduction} 
% G: best to avoid acronyms in titles

Let $\{\psi_k\}_{k=0}^{\rb-1}$ be an $\rb$-dimensional product basis
on $[-L,L]^2$ (cosine, Legendre, or Chebyshev products for example). % G edit
From the MC walk, accumulate outer products $\psi(\mathbf{r}_{g})\psi(\mathbf{r}_{g+1})^{\!\top}$
over all steps~$g$ to estimate the $\rb{\times}\rb$ Galerkin matrices
$\hat{A}_{jk}{=}\langle\psi_j,M^{(1)}\psi_k\rangle_\pi$,
$\hat{M}_{jk}{=}\langle\psi_j,\psi_k\rangle_\pi$, and tally vector
$\hat{h}_k{=}\langle\psi_k,h\rangle_\pi$.
The KKT-constrained resolvent removes the $\lam{=}1$ singularity from $\psi_0{=}1$:
\begin{subequations}\label{eq:mor}
\begin{align}
  &(\hat{M}-\hat{A})\hat{s} = \hat{A}\hat{q},\quad e_0^\top\hat{M}\hat{s}=0,\label{eq:kkt}\\
  &\tauint = 1 + \frac{2\,\hat{h}^\top\hat{s}}{\hat{h}^\top\hat{M}^{-1}\hat{h}}.\label{eq:taumor}
\end{align}
\end{subequations}

The cosine basis is the exact eigenbasis of this model; polynomial bases
(Legendre, Chebyshev) approximate the same eigenmodes with more pairs.

\subsection{Monte Carlo Algorithm}

$N_{\mathrm{batch}}$ independent batches are run.
Each batch simulates $G$ scattering steps of a single neutron; by ergodicity
this is equivalent to one generation of $G$ independent neutrons.
All methods share the same trajectory: each step $\mathbf{r}_g\!\to\!\mathbf{r}_{g+1}$
is tallied simultaneously by all estimators.
For the discrete chain, each step increments the cell-to-cell transition count
$P[i,j]$, where $i$ and $j$ are the cells of $\mathbf{r}_g$ and $\mathbf{r}_{g+1}$.
For Galerkin MOR, each step contributes an outer product to
$\hat{A}$, $\hat{M}$, and $\hat{h}$, accumulating to all matrix elements in one pass. % G edit
Each batch yields one estimate of $\tauint$ via Eq~\ref{eq:neumann} for discrete chain and Eq~\ref{eq:taumor} for MOR;  % using equation 7b ??? because we can't tally lamnda in one batch or ACCs
the $N_{\mathrm{batch}}$
independent estimates then provide a mean and standard uncertainty.

%%%%%%%%%%%%%%%%%%%%%%%%%%%%%%%%%%%%%%%%%%%%%%%%%%%%%%%%%%%%%%%%%%%%%
\section{Results}

\subsection{Setup}

All results use $\Sigma_t L = 33$, chosen to reproduce the dominant-mode
mixing rate of a large thermal reactor: the 3D diffusion approximation
gives $\lambda_1 = 1/\bigl(1+(\pi M/2L)^2\bigr) \approx 0.995$ for typical
migration area $M^2$ and core half-size $L$~\cite{PhdThesisJMiao}, matching
the scattering-chain eigenvalue $\lambda_{10} \approx 0.996$ at $\Sigma_t L{=}33$;
the resulting $\tauint^{\mathrm{ref}}{=}295$.
We use $\alpha = 0.5$ (tally $I=[-L/2,L/2]^2$) and $G = 10^7$ steps per batch.
The reference is $\tauint^{\mathrm{ref}} = 295.31$ from Eq.~\eqref{eq:tau},
retaining modes $m,n = 0,\ldots,500$ in each dimension and excluding the trivial
$(0,0)$ mode, leaving $501^2-1 = 251{,}000$ pairs (residual ${<}0.1\%$).
Accuracy runs use $N_{\mathrm{batch}}{=}40$; profiling runs use
$N_{\mathrm{batch}}{=}20$.
Hardware: $2{\times}$A100 GPUs, 8 CPU cores (Intel Xeon Gold 6248R).

\subsection{MOR Basis Comparison}
\begin{figure}[ht]
  \centering
  \includegraphics[width=\columnwidth]{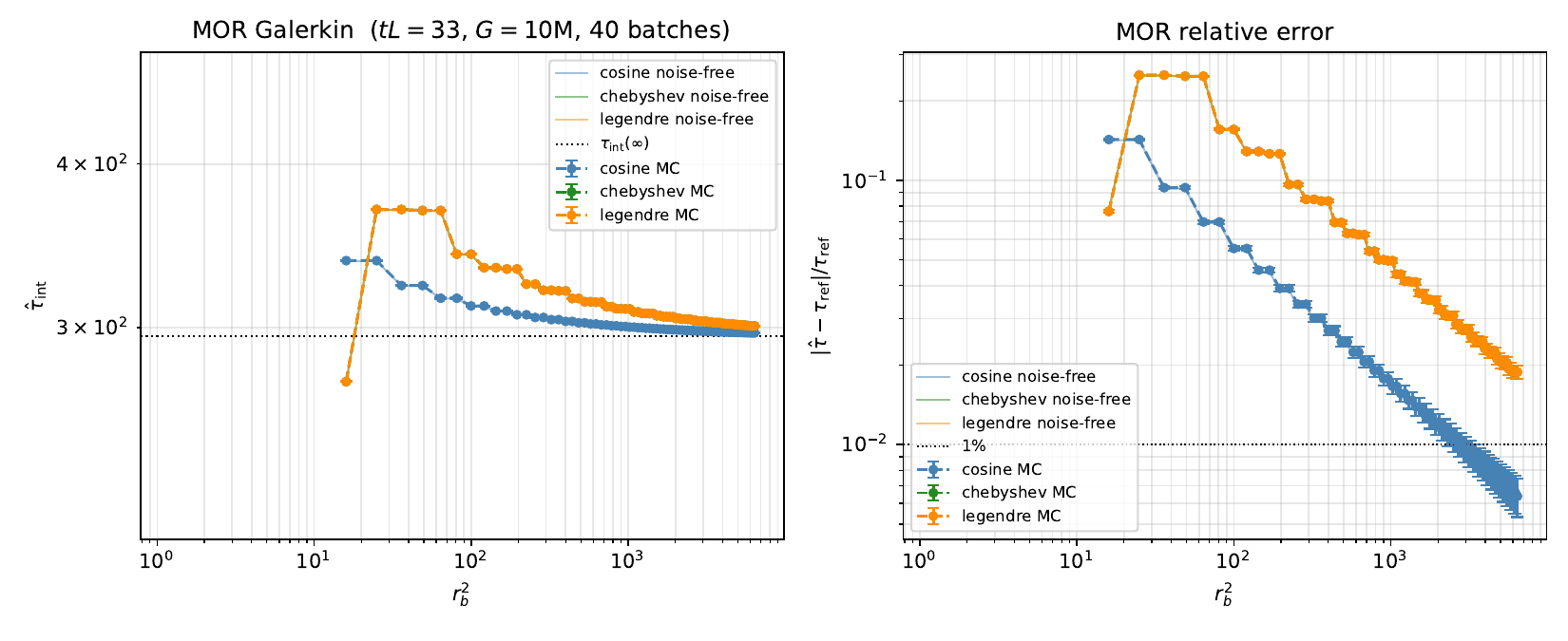}
  \caption{MOR accuracy vs.\ $\rb$ (total mode pairs) for three bases;
  NF lines = noise-free Galerkin; markers = MC ($N_{\mathrm{batch}}{=}40$, $G{=}10^7$).
  The cosine NF curve shows a staircase due to mode degeneracy (Fig.~\ref{fig:modes}).}
  \label{fig:mor}
\end{figure}

Three bases are compared across a range of $\rb$.
Fig.~\ref{fig:mor} shows $\tauint$ (left panel) and relative error
$\varepsilon{=}|\tauint^{\mathrm{MC}}{-}\tauint^{\mathrm{ref}}|/\tauint^{\mathrm{ref}}$
(right panel) as a function of total mode pairs $\rb$.
Solid lines are noise-free Galerkin (NF) results using exact quadrature;
markers with $\pm1\sigma$ bars are MC estimates.

The cosine basis (exact eigenbasis of this model) converges fastest:
$\rb{=}400$ ($20{\times}20$ tensor product) already achieves ${<}3\%$ error,
roughly $7{\times}$ fewer pairs than Legendre/Chebyshev, which require
$\rb{\approx}3025$ to reach comparable accuracy.
Legendre and Chebyshev product bases span the same polynomial subspace on
$[-L,L]^2$ and produce numerically identical NF and MC $\tauint$ at every $\rb$,
confirming that the result depends on the subspace, not the specific basis chosen.
MC estimates track the NF curves within statistical uncertainty at all bases
and orders, confirming that the Galerkin assembly from walk tallies is unbiased.

\subsection{Discrete $M$-Cell Convergence}
\begin{figure}[ht]
  \centering
  \includegraphics[width=\columnwidth]{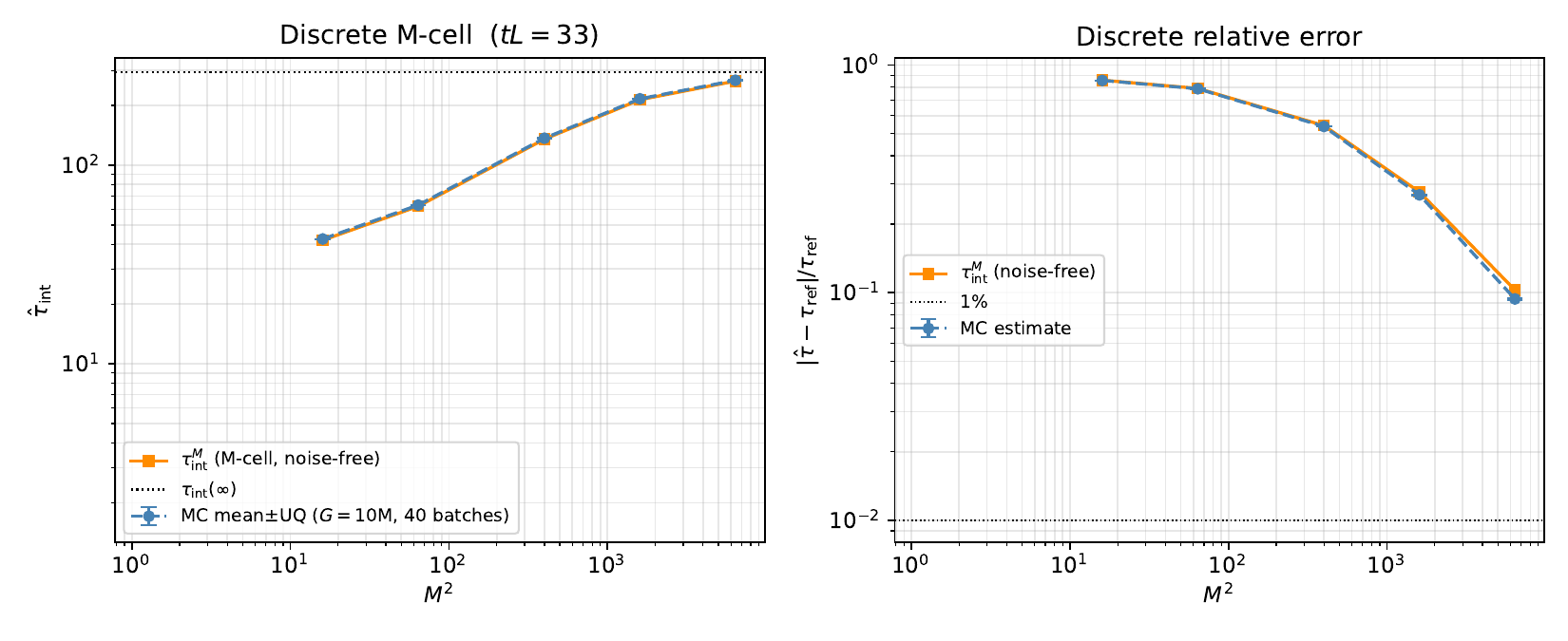}
  \caption{Discrete accuracy vs.\ $M$ (total cells).
  NF line = analytic $M$-cell chain; markers = MC tally-derived estimates.
  Even the NF chain is biased at finite $M$; MC adds statistical noise on top.}
  \label{fig:disc}
\end{figure}

The M-cell chain converges more slowly, with a discretization bias that persists at all finite $M$. 
Fig.~\ref{fig:disc} shows the analytic noise-free $M$-cell IACT
$\tauint^{(M)}$ (computed from the exact eigendecomposition restricted to
$M$ cells) alongside MC tally-derived estimates.

The discrete $M$-cell method carries an irreducible spatial discretization
bias: the finite-cell chain cannot resolve fine-scale spatial correlations
and systematically underestimates $\tauint$ at all finite $M$.
At $M{=}1600$ ($40{\times}40$) the NF bias is ${\sim}28\%$;
at $M{=}6400$ ($80{\times}80$) it is ${\sim}10\%$.
MC estimation adds additional noise on top of this bias.
In contrast, MOR with cosine $\rb{=}400$ achieves ${<}3\%$ error with no
mesh and no discretization bias.

\subsection{Efficiency}

\begin{figure}[ht]
  \centering
  \includegraphics[width=\columnwidth]{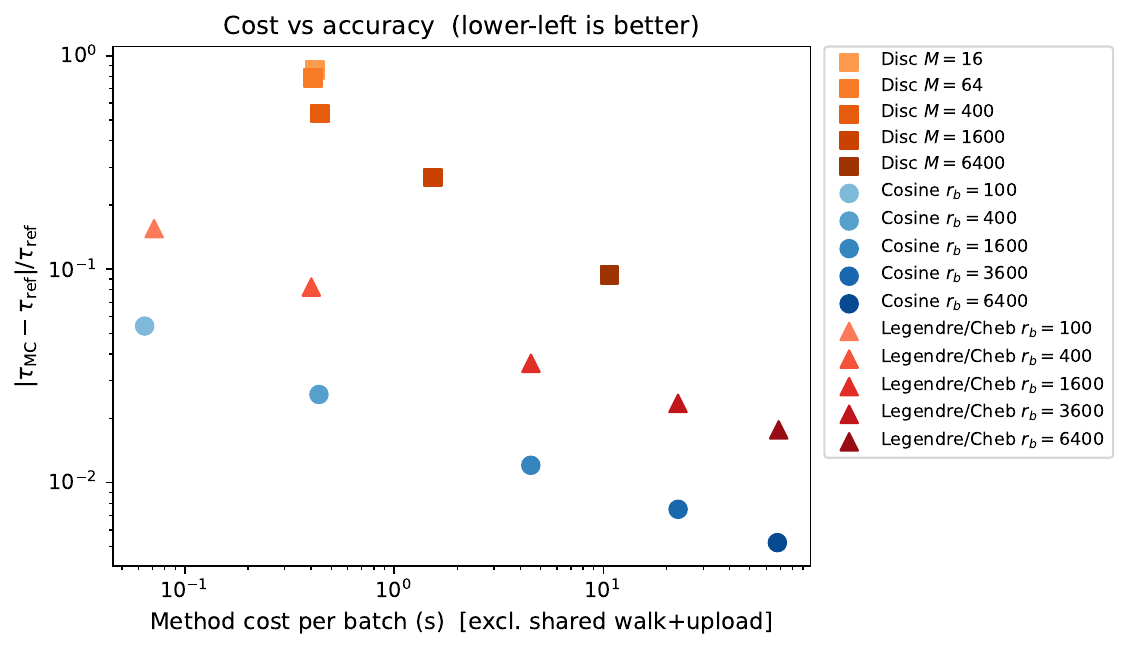}
  \caption{Efficiency scatter: relative error vs.\ per-batch method cost
  (shared walk+upload excluded).
  Squares = discrete; circles = cosine MOR;
  triangles = Legendre/Chebyshev MOR.
  Color encodes $M$ or $\rb$ (lighter = smaller).}
  \label{fig:eff}
\end{figure}

Fig.~\ref{fig:eff} plots relative error vs.\ per-batch method cost
(excluding the shared walk) for standalone profiling runs.
Points in the lower-left are preferred.

Table~\ref{tab:cost} breaks down per-batch costs.
For discrete, the dense $M{\times}M$ tally accumulation and Neumann
solve are reported together.
For MOR, assembly (GPU outer products) and the KKT solve (CPU dense linear
algebra) are reported separately.

\begin{table}[ht]
  \centering
  \caption{Per-batch costs, $G{=}10^7$, $2{\times}$A100 / 8 CPU cores
           (Intel Xeon Gold 6248R). MOR assembly is GPU-accelerated;
           discrete solve is CPU-only.}
  \label{tab:cost}
  \small
  \setlength{\tabcolsep}{4pt}
  \begin{tabular}{llrr}\toprule
    Method & Basis & Assembly & Solve \\ \midrule
    \multicolumn{4}{l}{\textit{Walk (shared): 7.7\,s/batch}} \\[2pt]
    Disc $M{=}400$   & -- & \multirow{3}{*}{(within solve)} & 0.44\,s \\
    Disc $M{=}1600$  & -- &                                  & 1.5\,s  \\
    Disc $M{=}6400$  & -- &                                  & 10.7\,s \\ \midrule
    \multirow{3}{*}{MOR $\rb{=}100$}
                     & Cosine & 0.06\,s &  2\,ms \\
                     & Chebyshev &0.07\,s &  1\,ms \\
                     & Legendre &0.07\,s &  1\,ms \\
\midrule                     
    \multirow{3}{*}{MOR $\rb{=}400$}
                     & Cosine & 0.42\,s & 16\,ms \\
                     & Chebyshev &0.39\,s & 20\,ms \\
                     & Legendre &0.39\,s &  9\,ms \\
\midrule                     
    \multirow{3}{*}{MOR $\rb{=}1600$}
                     & Cosine & 4.4\,s  & 97\,ms \\
                     & Chebyshev &4.4\,s  & 95\,ms \\
                     & Legendre &4.4\,s  & 98\,ms \\
\midrule                     
    \multirow{3}{*}{MOR $\rb{=}3600$}
                       & Cosine & 21.7\,s & 1.0\,s \\
                       & Chebyshev &21.8\,s & 0.86\,s \\
                       & Legendre &21.9\,s & 0.83\,s \\
\midrule                       
    \multirow{3}{*}{MOR $\rb{=}6400$}
                     & Cosine & 63.9\,s & 4.0\,s \\
                     & Chebyshev &63.8\,s & 4.5\,s \\
                     & Legendre &64.1\,s & 4.8\,s \\ \bottomrule
  \end{tabular}
\end{table}

MOR assembly scales as $O(\rb^2)$: each of the $G$ walk steps contributes
a rank-1 update to the $\rb{\times}\rb$ Galerkin matrices, and the GPU chunk
size must shrink as $\rb^{-1}$ to stay within memory, so per-batch cost
scales as $O(G\cdot\rb)$ for the GEMM, with the $\rb^{-1}$ chunking overhead
yielding $O(\rb^2)$ overall at fixed $G$.
The KKT solve is $O(\rb^{3})$ (dense factorization of the $\rb{\times}\rb$
system) but remains negligible until $\rb{\approx}3600$.
Discrete Neumann cost grows as $O(M^2)$: at $M{=}6400$ it already exceeds
$10$\,s with 8 CPU cores.

In a production run that already accumulates FET coefficients, the Galerkin
matrix assembly is part of the existing tally work regardless of whether
correlation prediction is performed.  The incremental cost of obtaining
$\tauint$ is therefore the KKT solve alone, making the solve cost the
appropriate basis for comparing methods.

%%%%%%%%%%%%%%%%%%%%%%%%%%%%%%%%%%%%%%%%%%%%%%%%%%%%%%%%%%%%%%%%%%%%%
\section{Conclusions}

Galerkin reduced-order modeling provides an efficient, unbiased estimate of the
integrated autocorrelation time $\tauint$ in batch Monte Carlo transport.

The exact 2D scattering chain benchmark shows that 
the same Markov operator can be represented either by discrete cell indicators or by a smooth reduced basis:
the discrete $M$-cell chain uses indicator functions over grid cells, while
Galerkin MOR uses smooth global bases; both converge to the exact $\tauint$
as $M,\rb\to\infty$, but at very different rates.
In a production run already accumulating FET coefficients, the assembly cost
is already paid; the incremental cost of $\tauint$ is the KKT solve alone,
making it the appropriate comparison metric.
On that basis, cosine MOR ($\rb{=}400$) achieves ${<}3\%$ error at a
16\,ms solve cost, while discrete $M{=}1600$ incurs ${\sim}28\%$
irreducible bias at 1.5\,s---nearly two orders of magnitude more
expensive for worse accuracy.
At equal solve cost (${\sim}1$\,s), cosine MOR at $\rb{=}3600$ reaches
${<}1\%$ error, while discrete $M{=}1600$ is stuck at ${\sim}28\%$ bias.

Future work will implement FET-style MOR tallies within the Multitype
Branching Process framework~\cite{Miao2016} for rigorous uncertainty
quantification in real criticality calculations, and extend the basis to
angle- and energy-dependent problems and heterogeneous geometries.

%%%%%%%%%%%%%%%%%%%%%%%%%%%%%%%%%%%%%%%%%%%%%%%%%%%%%%%%%%%%%%%%%%%%%
\section*{Acknowledgments}

The authors thank the Penn State Department of Nuclear Engineering for support and the
Institute for Computational and Data Sciences for HPC resources.

%%%%%%%%%%%%%%%%%%%%%%%%%%%%%%%%%%%%%%%%%%%%%%%%%%%%%%%%%%%%%%%%%%%%%
{\footnotesize
\bibliographystyle{ans}
\bibliography{mor_refs}
}

\end{document}